\documentclass[aps,prl,twocolumn,groupedaddress]{revtex4}
\usepackage{graphicx}
\usepackage{amsmath}

\begin{document}

\title{Virtual exceptional points in an electromechanical system}

\author{P. Renault}

\author{H. Yamaguchi}

\author{I. Mahboob}

\email{imran.mahboob@lab.ntt.co.jp}

\affiliation{NTT Basic Research Laboratories, NTT Corporation, Atsugi-shi, Kanagawa 243-0198, Japan}

\begin{abstract}
Non-Hermitian Hamiltonians can give rise to exceptional points (EPs) which have been extensively explored with nominally identical coupled resonators. Here a non-Hermitian electromechanical system is developed which hosts vibration modes that differ massively in their spatial profile, frequency and even the sign of their dissipation. An optomechanical-like parametric modulation is employed to dynamically couple 2 of these disparate modes and in the appropriate parameter regime their imaginary eigenvalues coalesce whilst their normal modes split. The presence of this {\it virtual} EP is confirmed via numerical simulations of the coupled equations of motions describing the dynamics of this non-degenerate system. These results suggest that virtual EPs can be accessed in systems composed from highly mismatched resonators whilst still maintaining access to the non-trivial phenomenon in their proximity, as recently demonstrated for topological operations, thus enabling non-Hermitian singularities to be more widely exploited with realistic physical systems. 
\end{abstract}

\maketitle

The Hamiltonians for non-conservative systems are non-Hermitian with finite imaginary eigenvalues and are generally regarded as being disadvantageous as information is lost from them. Counterintuitively it has been asserted that dissipation can actually be a resource if the parameters of a system can be exquisitely manipulated which leads to the emergence of EPs, also known as branch points or non-Hermitian degeneracies, where real and imaginary eigenvalues and the corresponding eigenvectors of the system coalesce \cite{NH1, NH2, NH3}. This concept has been explored in recent years most widely with coupled and nominally identical harmonic resonators leading to not only the observation of EPs but also to the emergence of non-trivial physics in their proximity for instance loss induced revival of lasing \cite{NH4, NH5, NH6, NH7, NH8, NH10, NH11}. Meanwhile a class of non-Hermiticity yielding only real eigenvalues near the EP was identified in systems with balanced gain and loss and labelled parity-time ($\mathcal{PT}$) symmetric which stimulated renewed interest in non-Hermitian systems \cite{bend1, bend3}. Experimentally $\mathcal{PT}$-symmetry has been frequently probed in ostensibly identical resonantly coupled electromagnetic resonators with balanced gain and loss which in addition to hosting EPs have also revealed a plethora of esoteric physics in their immediacy including non-reciprocity and $\mathcal{PT}$ lasers \cite{bend4, PT5, PT6, PT1, PT9, PT11}.

The exotic dynamics available in the vicinity of EPs makes their accessibility extremely desirable however the seemingly trivial requirement of the underlying harmonic resonators be identical proves formidably challenging to satisfy in practice due to the limitations in ultra-precise micro/nano fabrication \cite{roukX}. To address this challenge an optomechanical system \cite{cavopto} was recently harnessed to span the frequency gap between two highly non-degenerate mechanical vibration modes via a laser pumped optomechanical link \cite{NH12}. The resultant parametric coupling not only manifested a {\it virtual} EP but it also exhibited chirality in topological operations, a characteristic trait of EPs. Here an electromechanical system is developed which sustains 2 highly non-degenerate vibration modes (one with variable gain) which are parametrically coupled via a mechanical strain modulation at their frequency difference that results in the emergence of a tuneable virtual EP which not only can be discerned from direct experimental observations but is confirmed with numerical simulations indicating the imaginary eigenvalues converge in its proximity.                   

A simple model to illustrate the emergence of a virtual EP consists of two oscillators with natural frequencies $\omega_1 \ll \omega_2$ one with gain $\gamma_1<0$ and the other with loss $\gamma_2>0$ having amplitudes $x_1$ and $x_2$ respectively whose time dependence is depicted in Fig. 1(a). If these resonators can be  coupled with sufficient strength $\xi$ a nominally conservative configuration can be accessed where one now observes Rabi oscillations (namely coherent energy exchange) between the resonators, as depicted in Figs. 1(b). This transition from an open to a conservative system can mark the presence of an EP where the Rabi oscillations indicate the imaginary components of the constituent resonators have merged whilst their real components have undergone an avoided crossing \cite{NH6}. Naturally the ability to manipulate the coupling is key for this changeover to occur and in particular when $\omega_1 \neq \omega_2$ the frequency mismatch needs to be bridged.

\begin{figure}[!ht]
\includegraphics[scale=0.95]{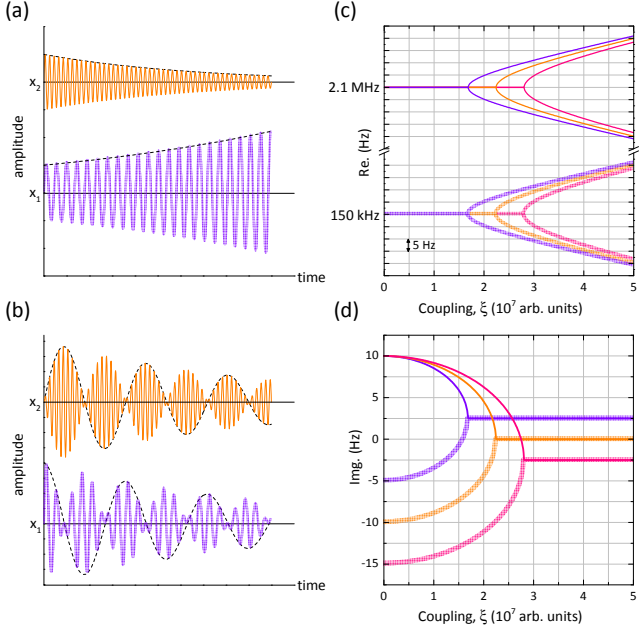}
\vspace{-0.5cm}
\small{\caption{(a) The time dependence of the both resonator's amplitude $x_1$, $x_2$ and frequency $\omega_1<\omega_2$ respectively which are offset for clarity, with the former (latter) captured by the thick faded violet (thin orange) line exhibiting gain (loss) corresponding to a negative (positive) imaginary component in the weak coupling limit ($\xi\approx 0$). (b) The time dependence of the resonators amplitude now in the strong coupling regime ($\xi\gg 0$) yields Rabi oscillations (black dashed lines) whose frequency is encapsulated by the real eigenvalues splitting. (c, d) The real and imaginary eigenvalues respectively as function of coupling strength extracted from equations 1 and 2, as detailed in ref. \cite{SM}, with $\omega_1=$150 kHz (thick faded lines), $\omega_2=$2.1 MHz (thin lines), $\gamma_2=20$ Hz, $\gamma_1=$-10 Hz (violet), -20 Hz (orange) and -30 Hz (pink). The coupling strength for which the imaginary eigenvalues coalesce is identical to where the real eigenvalues for both resonators split and this point marks the virtual EP which can be tuned by varying $\gamma_1$.}}
\end{figure}

To quantitatively examine this further, the equations of motion for these resonators are considered in terms of their acceleration, friction, restoring potential and coupling respectively given by

\vspace{-0.4cm}

\begin{eqnarray}
\ddot{x}_1 + \gamma_1\dot{x}_1 + \omega_1^2x_1 = x_2 \xi \cos\big((\omega_2-\omega_1)t\big)  \\
\ddot{x}_2 + \gamma_2\dot{x}_2 + \omega_2^2x_2 = x_1 \xi \cos\big((\omega_2-\omega_1)t\big) 
\end{eqnarray}

\noindent where the frequency mismatch between the resonators is mathematically compensated for by a time-dependent coupling at their frequency difference. These equations are analytically solved, as described in Supplemental Material \cite{SM}, to extract the real and imaginary eigenvalues the results of which are shown in Figs. 1(c) and 1(d). This analysis reveals that as the coupling strength is increased the imaginary eigenvalues of both resonators converge whilst concurrently their real eigenvalues, with a $\sim 2$ MHz frequency difference, split where this point marks the virtual EP and it quantitatively describes the transition from Fig. 1(a) to Fig. 1(b). A feature of this model can be discerned with a variable $\gamma_1<0$ which not only enables the resultant virtual EP to be tuned \cite{NH8, NH13} but in the strong coupling regime the sign of the coalesced imaginary eigenvalue can be adjusted from positive to negative yielding decaying or rising Rabi oscillations, with the former shown in Fig. 1(b), or even be entirely eliminated yielding a $\mathcal{PT}${\it-like} non-Hermitian system \cite{bend4}.

\begin{figure*}[!ht]
\vspace{6.8cm}
\includegraphics[scale=0.85]{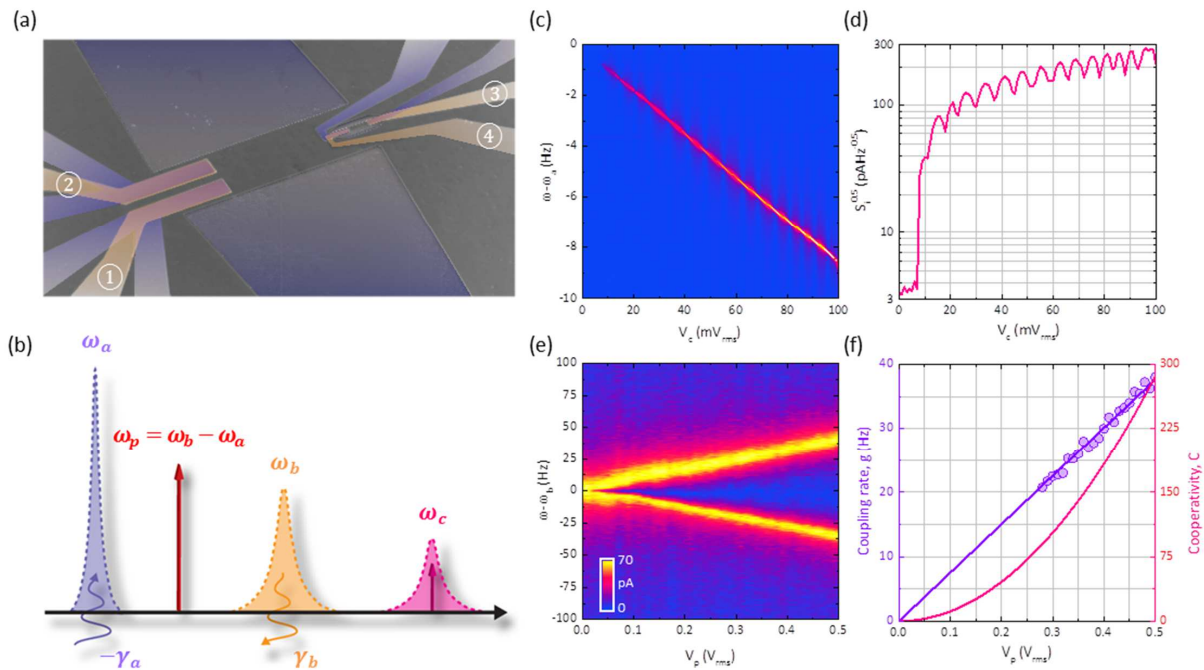}
\vspace{-7.0cm}
\small{\caption{(a) A false colour electron micrograph of the electromechanical system integrated with piezoelectric actuators, to transduce mechanical motion and to create dynamical strain, are composed from a two-dimensional electron gas located below the surface (violet) and gold Schottky electrodes (orange) sandwiching a layer of piezoelectric semiconductor \cite{SM}. (b) A spectral description of the electromechanical system where excitation of $\omega_c$ (pink arrow) generates gain in $\omega_a$ and excitation at $\omega_p$ (red arrow) creates parametric coupling between the non-degenerate modes $\omega_a$ and $\omega_b$. (c) The current noise spectral density around $\omega_a$, whilst $\omega_c$ is harmonically activated with voltage $V_c$, reveals the former undergoes self-oscillations which increase in amplitude as detailed in (d) and correspond to amplified gain as $\omega_c$ is more strongly excited. (e) The frequency response of $\omega_b$, under a weak harmonic probe of 4 mV$_{rms}$, as strain modulation at $\omega_p$ is activated with voltage $V_p$ that creates dynamic coupling between $\omega_a$ and $\omega_b$ which can be increased to the point where the two modes hybridize and undergo parametric normal-mode splitting. (f) The coupling rate (points) and the cooperativity (pink line), extracted from the peak splitting in (e) as a function of $V_p$, where the former reveals a linear dependence that is confirmed by a least-squares fit (violet line) and the latter indicates operation deep into the strong coupling regime namely ${\mathcal C} \gg 1$.}}
\end{figure*}

In order to experimentally investigate such a virtual EP, the electromechanical system shown in Fig. 2(a), and detailed elsewhere is employed \cite{imNL}. Specifically three vibrations modes are identified at frequencies $\omega_a/2\pi=154.277$ kHz, $\omega_b/2\pi=2.15236$ and $\omega_c/2\pi=2.5209$ MHz with damping rates $\gamma_a/2\pi=1$ Hz, $\gamma_b/2\pi=22$ Hz and $\gamma_c/2\pi=38.5$ Hz respectively whose spectral responses and mode profiles are described in ref. \cite{SM}. 

A unique characteristic of this system is the ability to create variable gain in $\omega_a$ with displacement $x_a$ by harmonically exciting $\omega_c$ which results in an intense mechanical vibration in the former as shown in Figs. 2(c) and 2(d) \cite{imNL2}. This behaviour can be captured by coupled Van der Pol equations of motion (equations 3 and 5 below) where the vibration $x_c$ creates a non-linear dissipation term $-\eta_a$ that neutralises $\gamma_a$ \cite{NL1, NL3, imNL2}. Indeed below a certain threshold vibration of $x_c$ only the miniscule thermo-mechanical fluctuations in $\omega_a$ exist \cite{cleland} but as the vibration $x_c$ is increased: $\gamma_a-\eta_a  \langle x_c^2 \rangle=0$ and $\omega_a$ self-oscillates through a Hopf bifurcation. Finally as $x_c$ becomes even larger, the self-oscillations in $\omega_a$ are amplified as: $\gamma_a-\eta_a  \langle x_c^2 \rangle < 0$ that is $\omega_a$ sustains tuneable gain \cite{imNL2}. 

The next challenge is to couple $\omega_a$ to one of the other modes in the system where clearly resonant coupling is unavailable as $\omega_a \neq \omega_b \neq \omega_c$. Consequently a phonon analogue of cavity-optomechanics is exploited to create parametric coupling between spectrally disparate modes \cite{imranK, cav1, cavopto}. Specifically the application of voltage $V$ in this system creates strain, from the piezoelectric effect, which enables the harmonic restoring potential of a mode to be modulated and thus its natural frequency to be tuned as $d\omega/dV$ \cite{imran1, cpl2}. Accordingly if the voltage $V_p$ induced strain modulation is activated at the frequency difference between two modes say $\omega_p=\omega_b$-$\omega_a$ then they can be parametrically coupled, in spite of their $\sim 2$ MHz frequency difference, via a classical analogue of the beam splitter interaction with coupling rate $g=V_p(\omega_p)\sqrt{\frac{d\omega_a}{dV}\frac{d\omega_b}{dV}}$ \cite{imranK, prx5, prx6}.

Based on this, $\omega_b$ is weakly harmonically probed whilst $V_p$ is activated at $\omega_p$, as shown in Fig. 2(e), and it results in an up-converted sideband from $\omega_a$ interfering with $\omega_b$ and undergoing parametric normal-mode splitting the characteristic feature of strongly coupled systems \cite{imranK}. The resultant parametrically enhanced coupling rate extracted from the peak splitting as $2g$ reveals a linear dependence on $V_p$ as shown in Fig. 2(f). The cooperativity ${\mathcal C}=4g^2/\gamma_a\gamma_b$ enables the potency of this coupling to be evaluated as a function of $V_p$, as shown in Fig. 2(f), which reveals a maximal ${\mathcal C}\approx 300$, the highest recorded in an all-mechanical setting.

With the establishment of parametric coupling between non-degenerate modes and the ability to create tuneable gain in one of them, the presence of virtual EPs as detailed in Figs. 1(c) and 1 (d) is explored. To that end the parametric coupling measured in Fig. 2(e) is repeated, but now with $\omega_a$ sustaining gain that is $\omega_c$ is also activated with $V_c=16$ mV$_{rms}$ so that $\omega_a$ self-oscillates just above threshold as described in Fig. 2(d) and this measurement configuration is summarised in Fig. 2(b). In addition to probing $\omega_b$ as a function of $V_p$, the spectral response of $\omega_a$ is now also simultaneously acquired and both outputs are shown in Figs. 3(a) and 3(b). The frequency response of $\omega_b$ now notably differs from the earlier case, in Fig. 2(e) where $\omega_a$ sustained loss, with the resonance splitting now not occurring at $V_p \rightarrow 0$ V$_{rms}$ and in addition a sharp peak also appearing. This peak corresponds to $\omega_a$ self-oscillating concurrently as confirmed in Fig. 3(b) which is up-converted into $\omega_b$. As the coupling between the modes is increased via $V_p$, the self-oscillation in $\omega_a$ extinguishes, in parallel a peak splitting in $\omega_b$ emerges where the former response reflects the imaginary eigenvalues (thick faded lines in Fig. 1(d)) and the latter the real eigenvalues (thin lines in Fig. 1(c)) of the electromechanical system thus suggesting these measurements can enable direct identification of the virtual EP. More strikingly, if this measurement is repeated with more gain in $\omega_a$, by setting $V_c=64$ mV$_{rms}$ as detailed in Fig. 2(d), it yields the outputs shown in Figs. 3(c) and 3(d) where now the peak splitting in $\omega_b$ is entirely eliminated with the self-oscillations in $\omega_a$ persisting up to the maximal value of $V_p$ which hints at the absence of a virtual EP. Note the intermediate spectral responses for $\omega_a$ and $\omega_b$ from $V_c=0$ to 64 mV$_{rms}$ are detailed ref. \cite{SM}.

These observations naively intimate that the virtual EPs can be extracted directly from the experimental measurements. However expiration of the self-oscillation in $\omega_a$ simply indicates the coupling voltage at which its negative imaginary eigenvalue becomes positive (thick faded violet line in Fig. 1(d)) marking the transition from gain to loss. This gauge can only yield the location of the virtual EP in the unlikely scenario the imaginary eigenvalue is entirely eliminated (thick faded orange line in Fig. 1(d)). On the other hand the onset of peak splitting $\Delta \omega_b$ in the spectral response of $\omega_b$ corresponds to the location of the virtual EP, as detailed in Figs. 1(c) and 1(d), but this point can only be reliably extracted if $\Delta \omega_b \ge \gamma_b$ as detailed in Fig. 3(e). Consequently the virtual EPs gleaned from this measure and located via $V_p$ are merely approximations, as the aforementioned inequality is only on the cusp of being satisfied, as shown in Fig. 3(f). 

Although the measurements described above superficially enable identification of the virtual EP, they provide no quantitative insight into the evolution of the real and imaginary eigenvalues of the electromechanical system as a function of coupling strength. Indeed the simple model in equations 1 and 2 is also insufficient for this purpose as it omits the non-linear effects at the heart of tuneable gain, parametric coupling and frequency pulling induced by $V_c$, $V_p$ and $V_b$. To that end a model is developed which is a combination of the coupled Van der Pol equations, that previously quantified the dynamics of the mode with gain \cite{imNL2}, and non-linear coupled equations, describing the parametric coupling between the non-degenerate mechanical modes \cite{imranK}, which results in the following three-mode equations of motion:

\begin{widetext}
\begin{eqnarray}
m_a \ddot{x}_a +(\gamma_a +\eta_a x_c^2)m_a \dot{x}_a +m_a \omega_a^2 x_a (1+\beta_a x_a^2+\Gamma_a x_c^2) = \lambda x_b \cos(\omega_pt)\\ 
m_b \ddot{x}_b +\gamma_b m_b \dot{x}_b +m_b \omega_b^2 x_b (1+\beta_b x_b^2) = \Lambda_b \cos\big((\omega_b +\delta_b)t\big)+\lambda x_a \cos(\omega_pt) \\ 
m_c \ddot{x}_c +(\gamma_c +\eta_c x_a^2)m_a \dot{x}_c +m_c \omega_c^2 x_c (1+\beta_c x_c^2+\Gamma_c x_a^2) = \Lambda_c \cos(\omega_c t) 
\end{eqnarray}
\end{widetext}

\noindent with mode masses $m_a \approx m_b \approx m_c$ determined from finite element method simulations shown in \cite{SM}. The parametric coupling between $\omega_a$ and $\omega_b$ is captured by the terms containing $\lambda \propto V_p$ and the harmonic driving of modes $\omega_b$ and $\omega_c$ is activated with amplitude $\Lambda_{b} \propto V_{b}, \Lambda_{c} \propto V_c$ respectively where the former can be detuned via $\delta_{b}$. The non-linear damping activated in $\omega_a$ is also accompanied by frequency dispersion as shown in Fig. 2(c) and this is accounted for by $\Gamma_a, \Gamma_c$ in the restoring potential in addition to the Duffing non-linearities $\beta_a, \beta_b, \beta_c$ which become active at large amplitudes. 

\begin{figure}[!ht]
\includegraphics[scale=0.95]{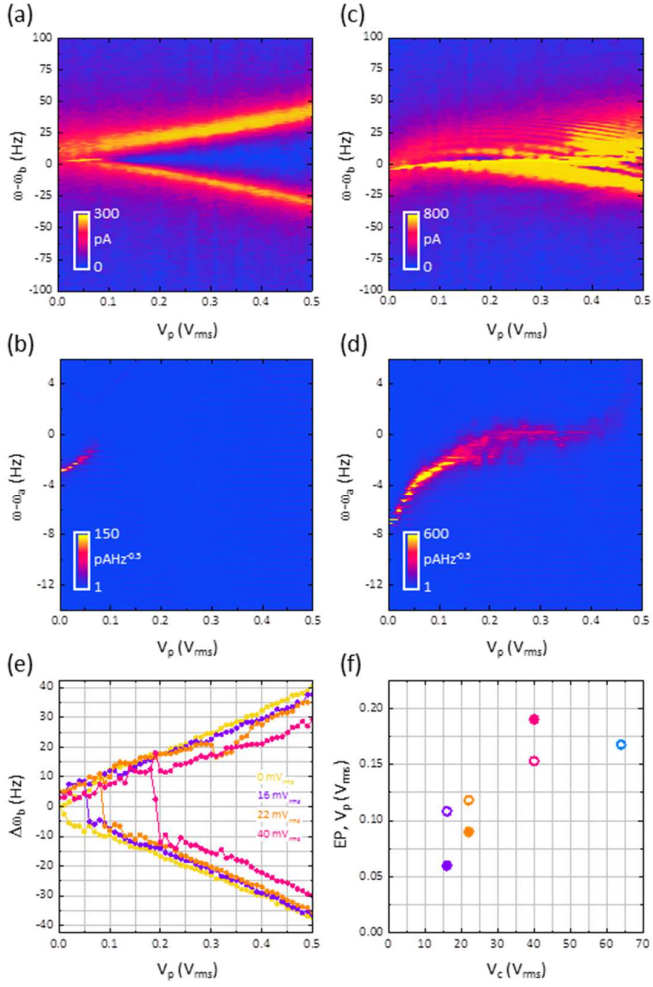}
\vspace{-0.5cm}
\small{\caption{(a, b) The simultaneously acquired spectral responses for $\omega_b$ and $\omega_a$ respectively as a function of parametric coupling strength $V_p$ between them, where the former is harmonically probed with a 4 mV$_{rms}$ excitation whilst the latter is interrogated via noise spectroscopy  with $V_c=16$ mV$_{rms}$ to create gain. (c, d) The same as (a, b) but now with $V_c=64$ mV$_{rms}$. (e) The peak splitting in $\omega_b$, for a range of colour coded $V_c$ values, extracted from the spectral responses in 2(e), (a) and \cite{SM} as a function of $V_p$. (f) The virtual EPs approximately extracted from the onset of peak splitting in $\omega_b$ (solid circle) in terms of $V_p$. Also shown are the EPs derived from the numerical simulations (open circles) which are broadly in-line with the experimental results.}}
\end{figure}

\begin{figure}[!hbt]
\includegraphics[scale=0.9]{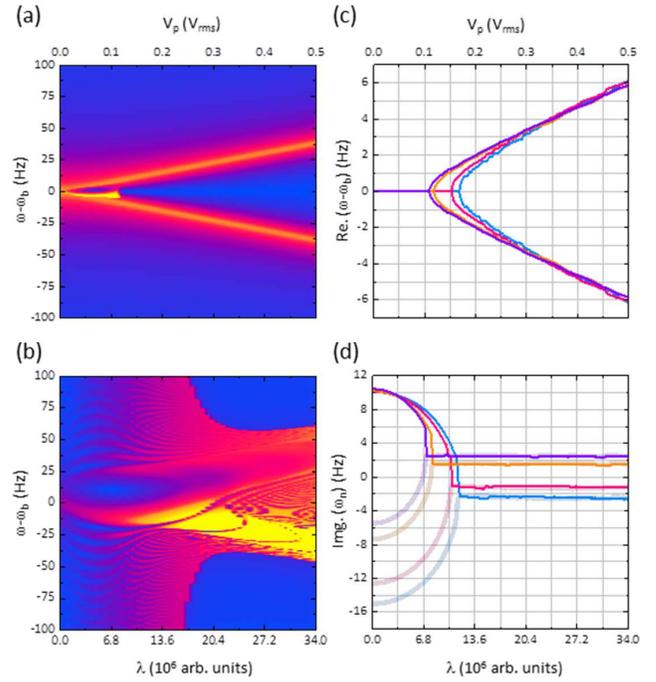}
\vspace{-0.0cm}
\small{\caption{(a, b) The numerically simulated response for $\omega_b$, derived from equations 3-5 as detailed in \cite{SM}, for the experiments depicted in Figs. 3(a) and 3(c) respectively. (c, d) The real and imaginary components extracted from equations 3-5, with the parameters determined from the spectral simulations in (a, b), by evaluating the time dependence of $x_a$ (thick faded lines) and $x_b$ (thin lines) where the real components correspond to the frequency of the Rabi oscillations gleaned from $\omega_b$ (where $\omega_a$ gives the same response) and the imaginary components are computed from the envelopes of $x_a$ and $x_b$. The imaginary eigenvalues also enable the gains in $\omega_a$ activated by $\omega_c$ to be deduced at $\lambda=0$ yielding $\gamma_a=-11, -15, -25, -30$ Hz corresponding to $V_c=16, 22, 40, 64$ mV$_{rms}$ (violet, orange, pink, cyan lines).}}
\end{figure}

Numerically solving equations 3-5 simultaneously for $\omega_b$, as detailed in \cite{SM}, with $V_c=16$ mV$_{rms}$ yields the result shown in Fig. 4(a) which reproduces the experimental response in Fig. 3(a). However as $V_c$ is increased in the experiment, the corresponding numerical simulations with increasing $\Lambda_c$ diverge at weak coupling namely $\lambda \rightarrow 0$ as shown in Fig. 4(b) but with stronger coupling the numerical simulations can capture all the details of the experiment including the asymmetric amplitude response, the absence of peak splitting and the interference fringes \cite{SM}. 

The parameters determined from the numerical simulations in Figs. 4(a) and 4(b) enable the time evolution of $x_a$ and $x_b$ to be evaluated, as schematically depicted in Figs. 1(a) and 1(b), from which the real eigenvalues for $\omega_b$ (where $\omega_a$ is identical except with $-\omega_p$ offset as shown in Fig. 1(c)) and the imaginary eigenvalues for both modes can be extracted resulting in Figs. 4(c) and 4(d) respectively. They in turn reveal the imaginary eigenvalues of both modes coalesce whilst simultaneously their real eigenvalues split, as the coupling between them is increased, thus enabling the virtual EPs to be identified in terms of $\lambda \propto V_p$ as shown in Fig. 3(f). Although the numerically simulated virtual EPs are broadly in-line with the experiments, the theoretical modelling is vital in aiding their accurate identification. The simulations also indicate that the sign of the coalesced imaginary eigenvalue can be adjusted as $V_c$ is varied and thus in the appropriate parameter range the imaginary eigenvalues can be entirely eliminated hence mimicking a $\mathcal{PT}$-like non-Hermitian system. 

Although a virtual EP can be approximately identified from the experiment and be located more precisely by the numerical simulation, further confirmation of its non-Hermitian nature would be achieved with an observation stemming exclusively from its presence. Indeed recent work with a virtual EP revealed such a {\it smoking-gun} with the observation of topological energy transfer between two non-degenerate mechanical modes \cite{NH12}. In the present work such topological encirculation of the virtual EPs proved challenging as the gain mode rapidly diverged as the system parameters where varied. However operating both modes with loss should still enable a virtual EP to condense \cite{NH12} with its encirculation expected to be technically more accessible and is the subject of a future study.

A virtual EP has been identified in an electromechanical system sustaining two highly dissimilar vibration modes from analytical modelling, experiments and numerical simulations. In contrast to previous work where EPs were invariably explored with nominally identical resonators, in the present case parametric coupling bridged their frequency mismatch enabling a virtual EP to emerge. Moreover by varying the gain in one of the modes enabled the virtual EP to be tuned with the sign of the imaginary eigenvalue of the coalesced system to be selectable or for it to even be entirely eliminated. Consequently these results pave way for the unique features of non-Hermitian singularities to be more easily accessed in a wider range of physical systems.

The authors are grateful to S. Miyashita for growing the heterostructure, A. Fujiwara and K. Nishiguchi for supplying the silicon cryo-amplifiers. This work was partially supported by MEXT KAKENHI Grant Number (JP15H05869).


\end{document}